\documentclass[aps,preprint,groupedaddress,a4]{revtex4-1}

\usepackage{amsmath,amssymb,epsfig}
\usepackage{braket}
\usepackage{slashed}

\begin{document}

\title{Classical Statistical simulation of Quantum Field Theory}

\author{Takayuki Hirayama}
\email[]{hirayama@isc.chubu.ac.jp}
\affiliation{Department of Mathematical and Physical Sciences, 
College of Science and Engineering, Chubu University
\\
1200 Matsumoto, Kasugai, Aichi, 487-8501, Japan}

\date{\today}

\begin{abstract}
We propose a procedure of computing the n-point function in perturbation theory of the quantum field theory as the average over the complex Gaussian noises in a classical theory. 
The complex Gaussian noises are the sources 
for the creation and annihilation of particles and the energy of the 
resultant configuration is the same as the zero point energy of the corresponding quantum field theory. 
\end{abstract}


\maketitle


\section{Introduction}

In quantum field theory, the vacuum state is the lowest energy state and 
its energy is the sum of $\hbar \omega/2$, the zero point energy in the 
free theory. The particles are continuously created and annihilated from 
the vacuum. These properties lead us to the idea that we mimic the 
quantum vacuum by the configuration in a classical theory with Gaussian random noises at each spacetime point which are the sources for the 
creation and annihilation of particles. Then it is easily computed that 
with the Gaussian noises, the configuration becomes non trivial and 
its energy is equal to 
the zero point energy in the quantum field theory. We will perturbatively 
show that the average of n-point function over the noises is exactly 
equal to the value of n-point function in quantum field theory. 

It has been known that in the non trivial background some of the quantum behaviors are realized in a classical theory. In%
~\cite{Boyer:1975re, Boyer:1975rf}%
, the each momentum modes of the electromagnetic fields have the 
amplitude with $\hbar$ and random phases. Then the motion of a 
charged particle in the non trivial electromagnetic fields shows some of quantum phenomena such as the blackbody radiation, the decrease of 
specific heats at low temperatures, and the absence of atomic collapse. In%
~\cite{Hirayama:2005ba, Hirayama:2005ac, Holdom:2006tt}%
, the random phase for each mode in the scalar field is introduced and then the zero point energy of the scalar field is realized in the classical theory. Then the loop Feynman graphs appear in the n-point function of scalar field, although not all the loop Feynman graphs in the quantum field theory appear. 

It has been also known that when the occupation number is large, the 
classical statistical approximation or the classical statistical Lattice 
simulation, is a good approximation to the quantum field theory%
~\cite{Aarts:1996qi, Aarts:1997kp}%
, and loop graphs appear. The initial conditions of the fields are chosen randomly and the fields are evolved by the classical equations of motion.
Then the expectation values of observables show the quantum effects, 
but it has been discussed that there is a discrepancy between the 
classical statistical approximation and quantum field theory.

Another attempt to realize the quantum theory from classical theory is discussed in%
~\cite{Wetterich:2010ts, Wetterich:2011zi}%
, where the wave function is introduced as the square of the probability distribution in a classical statistical model and the Schrodinger equation 
is derived.

The Gaussian noise is used in Langevin equation of stochastic quantization%
~\cite{Nelson:1966sp, Parisi:1980ys, Parisi:1983mgm, Klauder:1983sp, Damgaard:1987rr, Namiki:1993fd}%
.  In stochastic quantization, the quantum field theory is reached as the equilibrium limit of the system along the fictitious time direction. Recently the complex Langevin method has been applied to the theory with non-zero chemical potential and other theories in which the sign problem can not be avoided in Lattice simulation
~\cite{Aarts:2009uq, Aarts:2011ax, Anzaki:2014hba, Ichihara:2016uld, Aarts:2017vrv, Fujii:2017oti, Nagata:2018mkb, Scherzer:2018hid, Kogut:2019qmi, Scherzer:2019lrh}%
. There are other attempts and realizations of quantum mechanics or 
quantum field theory in terms of stochastic processes, such as 
stochastic variational method%
~\cite{Yasue1, Koide:2013vca, Kodama:2014cga}.

Compare to these methods, there is no fictitious time in our approach. 
But similar to the complex Langevin, we involve complexification of real 
fields. We explain our idea using a real scalar theory and give the main 
results in sec.\,\ref{sec2} and sec.\,\ref{sec2-2}, followed by the 
generalization to complex scalar, fermion and gauge fields in 
sec.\,\ref{sec3}. In sec.\,\ref{sec4}, we give summary.

\section{Complex noises}
\label{sec2}

In this section, we explain our idea of representing the creation and annihilation of particles from the vacuum by the Gaussian noises.
We first study a real scalar field in four dimensional Minkowski spacetime, and the generalization to the complex scalar, fermion and gauge fields will be discussed later.

In our construction, we complexify the real field $\phi(x)$ and introduce the complex Gaussian noises $J(x)$ at each spacetime point.
A complex Gaussian noise is constructed from two Gaussian noises, $J_r(x)$ and $J_i(x)$  as $J(x)=(J_r(x)+i J_i(x))/\sqrt{2}$ and $J^*(x)=(J_r(x)-i J_i(x))/\sqrt{2}$. The probability distributions, $P[J_r(x)]$ and $P[J_i(x)]$ for $J_r(x)$ and $J_i(x)$, are the Gaussian ones,
\begin{align}
  P[J_r(x)] &= \frac{1}{\sqrt{2\pi}\sigma}e^{-J^2_r(x)/2\sigma^2},
 \hspace{3ex}
 J_r(x) \in 	\mathbb{R},
  \\
  P[J_i(x)] &= \frac{1}{\sqrt{2\pi}\sigma}e^{-J_i^2(x)/2\sigma^2},
 \hspace{3ex}
 J_i(x) \in 	\mathbb{R},
\end{align}
where $\sigma$ is the standard deviation. Then the expectation value $\langle {\cal O}(x) \rangle$ is computed as the average over the noises,
\begin{align}
  \langle {\cal O}(x) \rangle =& \int \!\! {\cal D} J \,
  O(x)P[J],
  \\
  {\cal D}J =& \prod_y \int_{-\infty}^{\infty} d J_r(y)d J_i(y),
  \hspace{3ex}
  P[J] = \prod_y P[J_r(y)]P[J_i(y)].
\end{align}
Then we have,
\begin{align}
  \langle J(x) J(x') \rangle &= \langle J^*(x) J^*(x') \rangle
  = 0,
 \\
   \langle J(x) J^*(x') \rangle &=  \langle J^*(x) J(x') \rangle = \sigma^2 \delta^{(4)}(x-x'),
 \\
 \langle J^m(x) J^{*n}(x') \rangle &= \sigma^m m! \delta_{m,n}\delta^{(4)}(x-x').
\end{align}
We now study the solution of the equation of motion,
\begin{align}
 ( \Box + m^2 \!-\! i\epsilon) (\phi(x) + i J^*(x) ) = J(x),
\end{align}
where we introduce $i\epsilon$ in order to use the Feynman propagator
as a Green function.
The solution to this equation is written by using the Feynman propagator $D_F(x-y)$ as
\begin{align}
 \phi(x) &= \bigg[
 \frac{i}{\hbar} \int \!\! d^4y \, D_F(x-y)J(y)
 \bigg]
 -iJ^*(x).
\end{align}
The Feynman propagator satisfies
\begin{align}
 ( \Box + m^2 \! -\! i \epsilon ) D_F(x)=-i \hbar \delta^{(4)}(x).
\end{align}
Then the expectation value of $\phi(x)\phi(x')$, i.e. 2-point function,
is computed as 
\begin{align}
 \langle \phi(x) \phi_0(x') \rangle &= D_F(x-x'),
\end{align}
where we take $\sigma=\sqrt{\hbar/2}$.
In the same way, we compute 1-point function and 4-point function and they are
\begin{align}
 \langle \phi(x) \rangle =& \, 0,
\\
 \langle \phi(x_1) \cdots \phi(x_4) \rangle =&\, D_F(x_1-x_2) D_F(x_3-x_4)
 + D_F(x_1-x_3)D_F(x_2-x_4)
 \nonumber\\
 &+D_F(x_1-x_4)D_F(x_2-x_3).
\end{align}
These are equal to the expectation values in quantum field theory.

As mentioned in~\cite{Holdom:2006tt}, if the probability distributions differ from the Gaussian ones, the 4-point function has extra terms although 1- and 2-point functions are kept the same. Therefore we should use the Gaussian distribution and we can check the n-point function for any value of $n$ is equal to that in quantum field theory. Thus we have constructed a statistical theory which is equivalent with a free real scalar quantum field theory.
Since we have shown that the n-point function recovers the n-point function in quantum field theory, the expectation value of energy is the same as the zero point energy in quantum field theory. 
\begin{align}
 &\langle \int \!\! d^3x \, \frac{1}{2}\left( (\partial_t \phi(x) )^2 
 + (\partial_i \phi(x) )^2 + m^2 \phi^2(x) \right)
 \rangle
 =
 \hbar \int \!\! \frac{d^3k \, d^3x}{(2\pi)^3} \frac{k_0}{2}, 
\end{align}
where $k_0=+\sqrt{\vec{k}_i^2+m^2}$.

\section{Interacting theory}
\label{sec2-2}

\begin{figure}[thpb]
  \centering
 \fbox{\includegraphics[width=8.3cm]{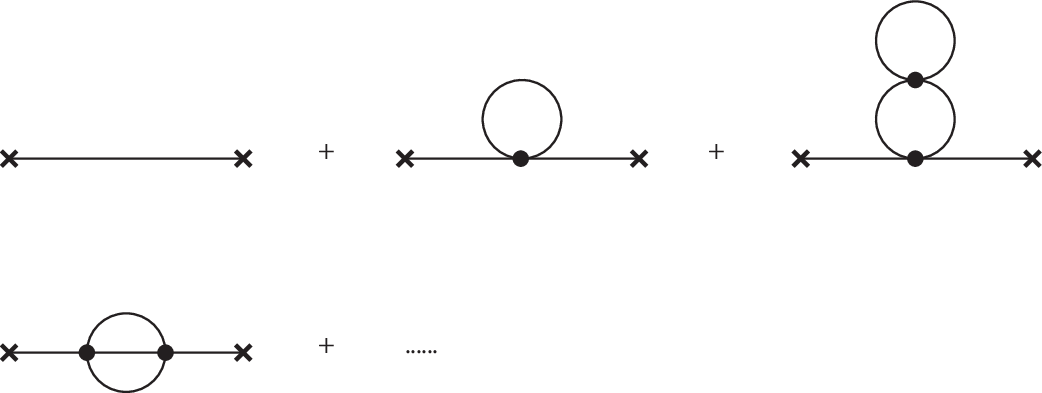}}
  \caption{Feynman graphs for $\phi^4$ theory}
  \label{t2}
\end{figure}

So far we have discussed the free theory. In this section, we study an interacting theory with the action,
\begin{align}
 S &= \!\!\int \!\! d^4x \, {\cal L}(\phi)
 \\
 &= \!\!\int \!\! d^4x
 \Big[ \frac{1}{2}(\partial_\mu\phi(x))^2 -\frac{1}{2}m^2\phi^2(x)+{\cal L}_{I}(\phi(x)) \Big].
\end{align} 
The interaction term ${\cal L}_{I}$ is  ${\cal L}_{I}(\phi(x))=-\lambda \phi^4(x)/4$ in the case of the $\phi^4$ theory for example. For this theory, we solve the following equation of motion,
\begin{align}
 ( \Box + m^2 \! -\! i\epsilon) (\phi(x) + i J^*(x) ) = J(x) +{\cal L}'_{I}(\phi(x)) 
  ,
\end{align}
where ${\cal L}'_{I}(\phi(x))=-\lambda\phi^3(x)$ for $\phi^4$ theory.
The solution for the equation is written
\begin{align}
 \phi(x) &= \phi_0(x) +\frac{i}{\hbar} \int \!\! d^4y \, D_F(x-y)
 {\cal L}'_{I}(\phi(y)),
 \\
 \phi_0(x) &= \bigg[
 \frac{i}{\hbar} \int \!\! d^4y \, D_F(x-y)J(y)
 \bigg]
 -iJ^*(x).
\end{align}
If we treat the interaction as a perturbation, the perturbative expansion of the solution is
\begin{align}
 \phi(x) =& \phi_0(x) +\phi_1(x) +\phi_2(x) +\phi_3(x) +\cdots,
\end{align}
and
\begin{align}
 \phi_1(x)=& \frac{i}{\hbar}\int \!\!d^4y \, D_F(x-y)
 {\cal L}'_{I}(\phi_0(y))
 ,
\ \\
 \phi_2(x)=& \frac{i}{\hbar}\int \!\! d^4y \, D_F(x-y)
 {\cal L}''_{I}(\phi_0(y))\phi_1(y)
 =\int \!\! d^4z \, \phi_1(z) \frac{\delta \phi_1(x)}{\delta \phi_0(z)}
 ,
 \\
 \phi_3(x)=& \frac{i}{\hbar}\int \!\! d^4y \, D_F(x-y)
 {\cal L}''_{I}(\phi_0(y))\phi_2(y)
 + \frac{i}{\hbar}\int \!\! d^4y \, D_F(x-y)
 {\cal L}'''_{I}(\phi_0(y))\phi_1(y)\phi_1(y)
 \\
 =&
 \int \!\! d^4z \, \phi_1(z) \frac{\delta \phi_2(x)}{\delta \phi_0(z)}
 .
\end{align}
In general we can express ($k\geq 1$)
\begin{align}
 \phi_k(x) \!=& \frac{i}{\hbar} \int \!\! d^4z \, d^4y \, \frac{\delta \phi_{k-1}(x)}{\delta \phi_0(z)} 
 D_F(z-y) {\cal L}'_{I}(\phi_0(y))
 \\
 =&\!
 \int \!\! d^4z \, \phi_1(z)
 \frac{\delta \phi_{k-1}(x)}{\delta \phi_0(z)}
 \label{31}
 \\
 =&  \left[ \int \!\! d^4z_1 \, \phi_1(z_1)
 \frac{\delta}{\delta \phi_0(z_1)}\right]\cdots
 \left[ \int \!\! d^4z_k \, \phi_1(z_k)
 \frac{\delta}{\delta \phi_0(z_k)}\right]\phi_0(x)
 .
\end{align}
We can check that $\phi_k(x)$ satisfies the equation of motion by the mathematical induction. From the equation \eqref{31}, 
\begin{align}
 ( \Box + m^2 \! -\! i\epsilon) \phi_k(x)
 =&\int \!\! d^4z \, \phi_1(z)
 \frac{\delta {\cal L'}^{(k-1)}_{I}(\phi(x))}{\delta \phi_0(z)}
 \\
 =&{\cal L'}^{(k)}_{I}(\phi(x)) ,
\end{align}
where  
${\cal L'}^{(k)}_{I}(\phi(x))$ is the $k$-th order terms of ${\cal L'}_{I}(\phi(x))$.
For the case of ${\cal L}'_I(\phi(x))=-\lambda \phi^{3}(x)$,
\begin{align}
{\cal L'}^{(k)}_{I}(\phi(x))
=\sum^{k}_{i_1=0}\sum_{i_2=0}^{k-i_1} -\lambda 
\phi_{i_1}(x) \phi_{i_2}(x)\phi_{k-i_1-i_2}(x).
\end{align}
We then try to compute the $n$-point function perturbatively. Using $\langle \phi_0(y_1) \phi_0(y_2) \rangle =D_F(y_1-y_2)$, we have
\begin{align}
  \langle \phi_1(x) \rangle &=
  \langle \frac{i}{\hbar}\int \!\!d^4y \, D_F(x-y)
 {\cal L}'_{I}(\phi_0(y)) \rangle
  \\
  &=
   \langle \phi_0(x) \frac{i}{\hbar}\int \!\!d^4y \,  {\cal L}_{I}(\phi_0(y)) \rangle
  -
  \langle \phi_0(x)\rangle \langle \frac{i}{\hbar}\int \!\!d^4y \,  {\cal L}_{I}(\phi_0(y)) \rangle
  ,
  \\
  \langle \phi_2(x) \rangle &= \langle \frac{i}{\hbar}\int \!\! d^4y \, D_F(x-y)
 {\cal L}''_{I}(\phi_0(y))\phi_1(y) \rangle 
  \\
  &=
  \langle \frac{i}{\hbar}\int \!\! d^4y_1 \, D_F(x-y_1)
 {\cal L}''_{I}(\phi_0(y_1)) \frac{i}{\hbar}\int \!\!d^4y_2 \, D_F(y_1-y_2)
 {\cal L}'_{I}(\phi_0(y_2)) \rangle
  \\
  &= \frac{1}{2}\langle \phi_0(x) 
  \frac{i}{\hbar}\int \!\!d^4y_1 \,  {\cal L}_{I}(\phi_0(y_1)) \frac{i}{\hbar}\int \!\!d^4y_2 \,  {\cal L}_{I}(\phi_0(y_2)) \rangle
  \nonumber
  \\
  &-\frac{1}{2}
  \langle \phi_0(x) \rangle\langle 
  \frac{i}{\hbar}\int \!\!d^4y_1 \,  {\cal L}_{I}(\phi_0(y_1)) \frac{i}{\hbar}\int \!\!d^4y_2 \,  {\cal L}_{I}(\phi_0(y_2)) \rangle
  \nonumber
  \\
  &-
  \langle \phi_0(x) 
  \frac{i}{\hbar}\int \!\!d^4y_1 \,  {\cal L}_{I}(\phi_0(y_1)) \rangle \langle\frac{i}{\hbar}\int \!\!d^4y_2 \,  {\cal L}_{I}(\phi_0(y_2)) \rangle.
\end{align}
Then we can show
\begin{align}
 \langle \phi(x) \rangle =
 \frac{
 \langle \phi_0(x) e^{\frac{i}{\hbar} \!\int \!\! d^4 z \, {\cal L}_{I}(\phi_0(z))} \rangle
 }{
 \langle e^{\frac{i}{\hbar} \!\int \!\! d^4 z \, {\cal L}_{I}(\phi_0(z))} \rangle
}.
\end{align}
Similiary, we can show
\begin{align}
 \langle \phi(x_1)\cdots \phi(x_n) \rangle =
 \frac{
 \langle \phi_0(x_1)\cdots \phi_0(x_n) 
 e^{\frac{i}{\hbar} \!\int \!\! d^4 z \, {\cal L}_{I}(\phi_0(z))} \rangle
 }{
 \langle e^{\frac{i}{\hbar} \!\int \!\! d^4 z \, {\cal L}_{I}(\phi_0(z))} \rangle
}.
\end{align}
For example, the Feynman graphs up to the second order for two point function in $\phi^4$ theory become FIG.\,\ref{t2}.

We notice that there are no bubble graphs and it is useful to define the generating functional
\begin{align}
  Z[\widetilde{J}] 
  =
 \frac{
 \langle e^{\frac{i}{\hbar} \!\int \!\! d^4 z \, {\cal L}_{I}(\phi_0(z))
 +\widetilde{J}(z)\phi_0(z)} \rangle
 }{
 \langle e^{\frac{i}{\hbar} \!\int \!\! d^4 z \, {\cal L}_{I}(\phi_0(z))} \rangle
}.
\end{align}
We come to compare $Z[\tilde{J}]$ with the generating functional 
$Z_{QFT}[\widetilde{J}]$ in quantum field theory,
\begin{align}
  Z_{QFT}[\widetilde{J}] 
  &= \frac{
 \int \!\! {\cal D}\phi \,
 e^{\frac{i}{\hbar}\int \!\! d^4 z \, {\cal L}(\phi) +\widetilde{J}(z)\phi(z)}
 }{
 \int \!\! {\cal D}\phi \,
 e^{\frac{i}{\hbar}\int \!\! d^4 z \, {\cal L}(\phi)}
 }.
\end{align}
In the perturbation theory, this becomes,
\begin{align}
  Z_{QFT}[\widetilde{J}] 
 &=\frac{
 \langle 
 e^{\frac{i}{\hbar}\int \!\! d^4 y \, {\cal L}_{I}(\phi(y)) +\widetilde{J}(y)\phi(y)}
 \rangle_0
 }{
 \langle 
 e^{\frac{i}{\hbar}\int \!\! d^4 z \, {\cal L}_{I}(\phi(z)) }
 \rangle_0
 },
\end{align}
where $\langle \cdots \rangle_0$ denotes that the contraction   
$\langle \phi(x)\phi(y) \rangle_0$ is done by $D_F(x-y)$ in the perturbation theory. In the previous section, we have shown that the expectation 
value of n-point function in our theory and that in quantum theory
are equal, and thus the expectation value of any function in both theories 
are equal. Therefore
\begin{align}
 Z[\widetilde{J}] = Z_{QFT}[\widetilde{J}].
\end{align}
Thus we have shown that our approach is equivalent with the quantum field theory in the perturbation level.

\section{complex scalar, fermion, and gauge field}
\label{sec3}

We can straightforwardly generalize our construction to the theory with complex scalar, fermion and gauge fields. From the previous sections,
once we can construct the Feynman propagators with using the 
Gaussian noises, the perturbation theory of our approach is equivalent with
the perturbation theory of quantum field theory.

A free complex scalar field $\Phi(x)$ is composed with two real scalar fields, $\phi_1(x)$ and $\phi_2(x)$, i.e. $\Phi(x)=(\phi_1(x)+i \phi_2(x))/\sqrt{2}$, and we introduce two complex Gaussian noises, $J_1(x)$ and $J_2(x)$. We take 
\begin{align}
 \phi_1(x) &= \left[
 \frac{i}{\hbar} \int \!\! d^4y \, D_F(x-y)J_1(y)\right]-iJ^*_1(x),
 \\
  \phi_2(x) &= \left[
  \frac{i}{\hbar} \int \!\! d^4y \, D_F(x-y)J_2(y)\right]-i J^*_2(x).
\end{align}
Since $\Phi^*(x)$ is the complex conjugate of $\Phi(x)$,  we may naively take $\Phi^*(x)=(\phi^*_1(x) -i \phi^*_2(x))/\sqrt{2}$. However this does not recover the Feynman propagator, but we should take 
$\Phi^*(x)=(\phi_1(x) -i \phi_2(x))/\sqrt{2}$. Then,
\begin{align}
 &\Phi(x) = \left[\frac{i}{\hbar} \int \!\! d^4y \, D_F(x-y)J(y)\right]
 -i \bar{J}^*(x),
 \label{c1}
 \\
 &\Phi^*(x) = \left[\frac{i}{\hbar} \int \!\! d^4y \, D_F(x-y)\bar{J}(y)\right]
 -iJ^*(x),
 \label{c2}
 \\
 &J(x) = \frac{J_1(x) +i J_2(x)}{\sqrt{2}},~~ 
 \bar{J}^*(x) = \frac{J_1^*(x) +i J_2^*(x)}{\sqrt{2}},
 \\
 &\bar{J}(x) = \frac{J_1(x) -i J_2(x)}{\sqrt{2}},~~
 J^*(x) = \frac{J_1^*(x) -i J_2^*(x)}{\sqrt{2}}.
\end{align}
Taking the average, we obtain
\begin{align}
  &\langle \Phi(x)\Phi^*(x')\rangle=  D_F(x-x'),
 \\
  &\langle \Phi(x)\Phi(x')\rangle
  =
  \langle \Phi^*(x)\Phi^*(x')\rangle =0.
\end{align}
For the interacting theory, the equations which $\Phi(x)$ and $\Phi^*(x)$
should satisfy are,
\begin{align}
 &\!\!\!(\Box +m^2-i\epsilon) (\Phi(x) +i\bar{J}^*(x) )
  = J(x) +\frac{\partial {\cal L}_{I}(\Phi^*\Phi)}{\partial \Phi^*(x)},
 \\ 
 &\!\!\!\!(\Box +m^2-i\epsilon) (\Phi^*(x) +iJ^*(x) )
 = \bar{J}(x) +\frac{\partial {\cal L}_{I}(\Phi^*\Phi)}{\partial \Phi(x)}.
\end{align}
We take \eqref{c1} and \eqref{c2} as the zero-th order solutions and solve these equations perturbatically. Then as in the previous section, the n-point function in our approach is the same as that in quantum field theory.

For a Dirac fermion field $\psi_\alpha(x)$, we introduce complex Gaussian distributions $J_{1\alpha}(x)$ and $J_{2\alpha}(x)$ and further introduce the Grassmann numbers, $\theta_\alpha(x)$ and the conjugate $\bar{\theta}_\alpha(x)$ in order to realize the anti-commutation. We then combine these into $J_{\alpha}(x)$ and $\bar{J}_\alpha(x)$,
\begin{align}
 J_\alpha(x) &= \frac{J_{1\alpha}(x) \!+\! i J_{2\alpha}(x)}{\sqrt{2}} \theta_\alpha(x)
 ,&
 \bar{J}^*_\alpha(x) &= \frac{J^*_{1\alpha}(x) \!+\! i J^*_{2\alpha}(x)}{\sqrt{2}} \theta_\alpha(x)
 ,
 \\
 \bar{J}_\alpha(x) &= \frac{J_{1\alpha}(x) \!-\! i J_{2\alpha}(x)}{\sqrt{2}} \bar{\theta}_\alpha(x)
 ,&
 J^*_\alpha(x) &= \frac{J^*_{1\alpha}(x) \!-\! i J^*_{2\alpha}(x)}{\sqrt{2}} \bar{\theta}_\alpha(x)
 ,
\end{align}
where we do not take summation over the spin index $\alpha$.
We define $\psi_\alpha(x)$ and $\bar{\psi}_\alpha(x)$ as
\begin{align}
  \psi_\alpha (x) &=
  \left[\frac{i}{\hbar}\int \!\!d^4y \,
  S_{F\alpha\gamma}(x-y) J_\gamma(y) \right]
  -i\bar{J}_\alpha^*(x)
  ,
  \\
  \bar{\psi}_\alpha (x) &=\left[
  \frac{i}{\hbar}\int \!\!d^4y \,
  \bar{J}_{\gamma}(y)S_{F\gamma\alpha}(y-x)\right]
  -i J_\alpha^*(x)
  ,
\end{align}
where 
$S_{F\alpha\beta}(x-x')$ is the Feynman propagator for a Dirac fermion,
\begin{align}
 S_{F\alpha\beta}(x-x')
 &=
 \hbar \int \frac{d^4k}{(2\pi)^4}
 \frac{i(\slashed{k}+m)_{\alpha\beta}e^{-ik(x-x')}}{k^2-m^2+i\epsilon}.
\end{align}
We now have the Grassmann numbers $\theta_\alpha(x)$ and $\bar{\theta}_\alpha(x)$, thus
the computation of expectation values should be modified, 
\begin{align}
  \langle {\cal O}(x) \rangle =& \int \!\! {\cal D} J {\cal D}\theta \,
  O(x)P[J] P[\theta],
  \\
  &{\cal D}\theta = \prod_{y,\alpha} \int \!\! d \theta_\alpha(y)
  d \bar{\theta}_\alpha(y),
  \hspace{3ex}
  P[\theta] = \prod_{y,\alpha} e^{\theta_\alpha(y)}e^{\bar{\theta}_\alpha(y)}.
\end{align}
Then we compute the expectation values and obtain
\begin{align}
  \langle \psi_\alpha(x)\bar{\psi}_\beta(x')\rangle
 &=  S_{F\alpha\beta}(x-x'),
 \\
  \langle \psi_\alpha(x)\psi_\beta(x')\rangle &=
  \langle \bar{\psi}_\alpha(x)\bar{\psi}_\beta(x')\rangle=0.
\end{align}
In the interacting theory, the Dirac fermion should satisfy
\begin{align}
 &
 (i\gamma^\mu \partial_\mu -m )_{\alpha\beta} 
 (\psi_{\beta}(x) +i \bar{J}_{\beta}^*(x) )
 = J_\alpha(x) +\frac{\partial {\cal L}_{I}}{\partial \bar{\psi}_\alpha(x)},
 \\
 &\partial_\mu(\bar{\psi}_\alpha(x) +iJ_\alpha^*(x)) 
  (-i\gamma^\mu)_{\alpha\beta} 
 -(\bar{\psi}_\alpha(x) +iJ^*(x)_\alpha) m_{\alpha\beta}
 = \bar{J}_\beta(x) +\frac{\partial {\cal L}_{I}}{\partial \psi_\beta(x)}
 .
\end{align}

For a gauge field $A_\mu^a$, we start with the action with the ghost $c^a$ and $\bar{c}^a$,
\begin{align}
 {\cal L} &= -\frac{1}{4} F^a_{\mu\nu} F^{a\mu\nu}  -\frac{1}{2\alpha}(\partial^\mu A_\mu^a)^2
 + i\bar{c}^a \partial^\mu D^{ab}_\mu c^b,
\end{align}
where $a$ is the gauge index and $\alpha$ is the gauge fixing parameter.
We take
the gauge field, 
\begin{align}
 A^a_\mu(x) &=\left[ 
  \frac{i}{\hbar}\int \!\! d^4y \,
  D^{ab}_{F\mu\nu}(x-y) g^{\nu\rho}J^{b}_\rho (y) \right]
  -i \bar{J}^{a}_\mu (x),
\end{align}
where $J^a_\mu(x)$ is a complex Gaussian distribution and 
$\bar{J}^{a}_\mu(x)=\sum_\rho g_{\mu\rho} (J^a_\rho(x))^*$ 
so that
$\langle J^a_\mu(x)\bar{J}^b_\nu(y)\rangle = 
g_{\mu\nu}\delta^{ab}\delta^{(4)}(x-y)$. $J^a_\mu(x)$ transforms as adjoint and vector under the gauge and Lorentz transformations.
The expectation value of 2-point function then recovers the Feynman propagator,
\begin{align}
 \langle A^a_\mu(x)A^b_\nu(y)\rangle &=
 D^{ab}_{F\mu\nu}(x-y).
\end{align}
The $c^a(x)$ and $\bar{c}^a(x)$ fields take
\begin{align}
 &c^a(x) = \left\{
  \left[\frac{i}{\hbar}\int \!\!d^4y \,
  D^{ab}_{c}(x-y) J^b(y)  \right]
  -i \bar{J}^{a*}(x) \right\}\theta^a(x),
 \\
 &\bar{c}^a(x) = \left\{ 
  \left[\frac{i}{\hbar}\int \!\!d^4y \,
  D^{ab}_{c}(x-y) \bar{J}^b(y) \right] 
  -i J^{a*}(x) \right\} \bar{\theta}^a(x),
\end{align}
where the summation is not taken over the index $a$, but is taken over the index $b$, $D_c^{ab}(x-y)$ is the Feynman propagator, $J^a(x)$ is a complex Gaussian distribution and $\theta^a(x)$ and $\bar{\theta}^a(x)$ are real Grassmann numbers. We obtain
\begin{align}
 &\langle c^a(x)\bar{c}^b(x')\rangle =
 D^{ab}_{c}(x-x').
\end{align}
We thus have recovered the Feynman propagators for complex scalar field, Dirac fermion, and gauge fields using Gaussian distributions.


\section{Summary and Discussion}
\label{sec4}

We discussed a statistical theory which can compute the n-point function in quantum field theory in the perturbative level. The Gaussian noises which are introduced in our procedure are also needed in the stochasitic quantization~\cite{Nelson:1966sp, Parisi:1980ys, Parisi:1983mgm, Klauder:1983sp, Damgaard:1987rr, Namiki:1993fd}%
. The difference from the stochastic quantization is that we do not need the fictitious time, but the similarity to the complex Langevin method is that we involve the complexification of the fields.

In quantum field theory, the perturbative expansion of qunatum field theory
suffers from various UV and/or IR divergences and the appropriate regularization methods, such as the UV cutoff and the dimensional regularization, and the renomalization are needed.
In our approach, we can apply the same regularization and renormalization 
used in quantum field theory.

\begin{acknowledgments}
The author would like to thank Bob Holdom and Feng-Li Lin for discussion.
\end{acknowledgments}

\end{document}